\documentclass[a4paper]{iopconfser}
\usepackage{amsmath,amssymb,amsfonts,mathtools,hyperref}

\usepackage[document]{ragged2e}
\usepackage[english]{babel}
\usepackage{graphicx}
\usepackage{url}
\usepackage{color}
\usepackage{listings}
\usepackage{textcomp}
\usepackage{endnotes}
\usepackage{multirow}
\usepackage{float}
\usepackage{array}
\usepackage{multicol}

\begin{document}

\title{The QES sextic and Morse potentials: exact WKB condition and supersymmetry}

\author{Alonso Contreras-Astorga$^{1}$ and A. M. Escobar-Ruiz$^{2}$}

\affil{$^1$CONAHCYT- Physics Department, Cinvestav, P.O. Box. 14-740, 07000, Mexico City, Mexico}\vspace{3mm}
\affil{$^2$Departamento de F\'{i}sica, Universidad Aut\'onoma Metropolitana Unidad Iztapalapa, San Rafael Atlixco 186, 09340 CDMX, M\'exico}

\email{alonso.contreras@cinvestav.mx; admau@xanum.uam.mx}

\begin{abstract}\justifying
In this paper, as a continuation of [Contreras-Astorga A., Escobar-Ruiz A. M. and Linares R., \textit{Phys. Scr.} {\bf99} 025223 (2024)] the one-dimensional quasi-exactly solvable (QES) sextic potential $V^{\rm(qes)}(x) = \frac{1}{2}(\nu\, x^{6} +  2\, \nu\, \mu\,x^{4} +  \left[\mu^2-(4N+3)\nu  \right]\, x^{2})$ is considered. In the cases $N=0,\frac{1}{4},\,\frac{1}{2},\,\frac{7}{10}$ the WKB correction $\gamma=\gamma(N,n)$ is calculated for the first lowest 50 states $n\in [0,\,50]$ using highly accurate data obtained by the Lagrange Mesh Method. Closed analytical approximations for both $\gamma$ and the energy $E=E(N,n)$ of the system are constructed. They provide a reasonably relative accuracy $|\Delta|$ with upper bound $\lesssim 10^{-3}$ for all the values of $(N,n)$ studied. Also, it is shown that the QES Morse potential is shape invariant characterized by a hidden $\mathfrak{sl}_2(\mathbb{R})$ Lie algebra and vanishing WKB correction $\gamma=0$.  
\end{abstract}

\section{Introduction}

\justifying

Quasi-exactly solvable (QES) systems in quantum mechanics represent a class of Hamiltonian operators for which solely a finite sector of the whole spectra can be determined exactly (closed analytical expressions for spectra and wavefunctions) using algebraic methods \cite{Turbiner1988QuasiexactlysolvablePA}. Unlike exactly solvable systems (such as hydrogen atom, the harmonic oscillator and the Morse potentials), in QES systems only a subset of eigenstates and eigenvalues can be found exactly, while the rest remains unknown. 

QES systems play an intermediate role between exactly solvable models and non-integrable systems. They are instrumental in understanding the boundaries of solvability in quantum mechanics and have applications in various fields, such as mathematical physics, quantum field theory, and condensed matter physics.

In a recent publication \cite{Escobar2024}, the present authors studied the SUSY partner Hamiltonians of the QES sextic potential\footnote{There is a global factor of $\frac{1}{2}$ with respect to the definition given in \cite{Escobar2024}} $V^{\rm(qes)} = \frac{1}{2}(\nu\, x^{6} +  2\, \nu\, \mu\,x^{4} +  \left[\mu^2-(4N+3)\nu  \right]\, x^{2})$ from an algebraic perspective. If $N \in \mathbb{Z}^+$, the potential $V^{\rm(qes)}$ admits $(N+1)$ exact solutions (including the ground state) described by the underlying hidden $\mathfrak{sl}_2(\mathbb{R})$ Lie algebra. Hence, for each $N \in \mathbb{Z}^+$ the SUSY partner potential $V_1(x,N)$ can be constructed \cite{Gangopadhyaya1995}. Interestingly, in the variable $x$ or $z=x^2$, it was found that the corresponding SUSY Hamiltonian with potential $V_1(x,N)$ does not possess a hidden $\mathfrak{sl}_2(\mathbb{R})$ algebraic structure.    

In this work, we aim to construct closed compact analytical approximations for the energy and for the so called WKB correction $\gamma$ of the QES sextic potential. The SUSY partner Hamiltonians of the QES Morse potential are also investigated at the level of the corresponding algebraic operators governing the polynomial part of the exact QES eigenfunctions.


\section{Exact WKB condition: analytical interpolations}
\label{sec WKB}
\bigskip

Let us consider the one-dimensional Schr\"odinger  equation for the QES sextic potential

\begin{equation}\label{Vred}
{\cal H}_{\rm qes}\,\psi(x)\ = \  \bigg[-\frac{1}{2}\frac{d^2}{dx^2}  \ + \ \frac{1}{2}\,\big(x^{6}\ + \ 2\, x^{4}\ - \ 2\,(2N+1) \, x^{2}\,\big)\bigg]\,\psi(x) \ = \ E\,\psi(x)  \ ,
\end{equation}
($\hbar=1$, $m=1$) where $N$ is a real parameter. The problem is defined in the domain $x \in (-\infty,\infty)$ with $\psi \in {\cal L}^2$.
For any $N$, using the Lagrange Mesh Mathematica Package \cite{JCR} (LMMP), the numerical eigenfunctions $\psi_n(x)$ and eigenvalues $E_n$ can be computed in a straightforward manner. Since the LMMP allows us to achieve very accurate numerical results with high precision (not less than 15 significant figures in accuracy) we can denote them, in practice, as the \textit{exact} ones. These exact values serve as the starting point to study the so called exact WKB condition (see below).

\vspace{0.1cm}

The Bohr-Sommerfeld quantization condition combined with the {\it WKB correction} $\gamma=\gamma(N,n)$ (arising from the sum of higher-order WKB terms taken at the exact energies) define the exact WKB condition \cite{JCRBS} which, by construction, must reproduce the \textit{exact} energies. Explicitly, it is described by the equation

\begin{equation}
\label{BSCon}
\int_{x_1(E)}^{x_2(E)}\sqrt{2\,(\,E \ - \ V^{\rm(qes)}(x,N)\,)\,dx} \ = \ \pi\,\bigg( n \,+\,\frac{1}{2}\,+\,\gamma\, \bigg) \ ,
\end{equation}
where $V^{\rm(qes)}(x,N)$ is the sextic potential function appearing in (\ref{Vred}), $x_1(E)$ and $x_2(E)$ are the corresponding (symmetric) turning points obeying $V(x)=E$, $n\in \mathbb{Z}^+$ is the principal quantum number labeling the state (the number of nodes in $\psi$), and $E=E_n$ is taken as the \textit{exact} energy obtained from the LMMP. Therefore, by fixing $(N,n)$, one can consider (\ref{BSCon}) as the defining equation for the WKB correction $\gamma=\gamma(N,n)$.

\vspace{0.1cm}

\begin{figure}[h]
\centering
\includegraphics[width=7.3cm]{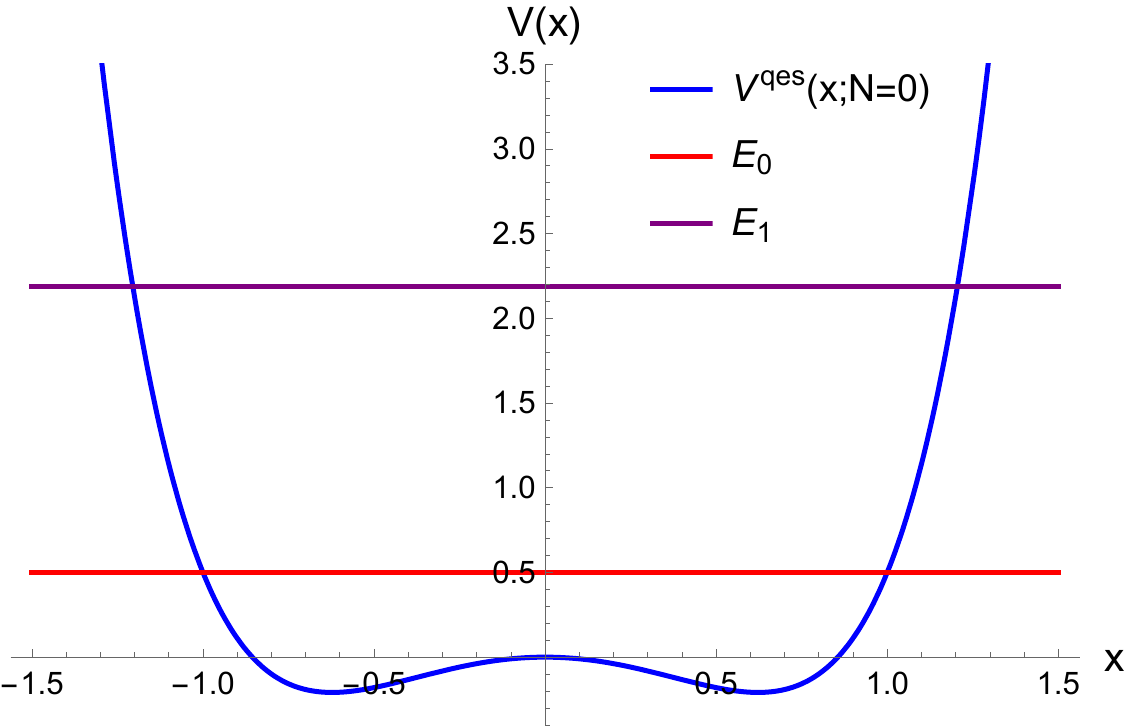} \qquad \includegraphics[width=7.3cm]{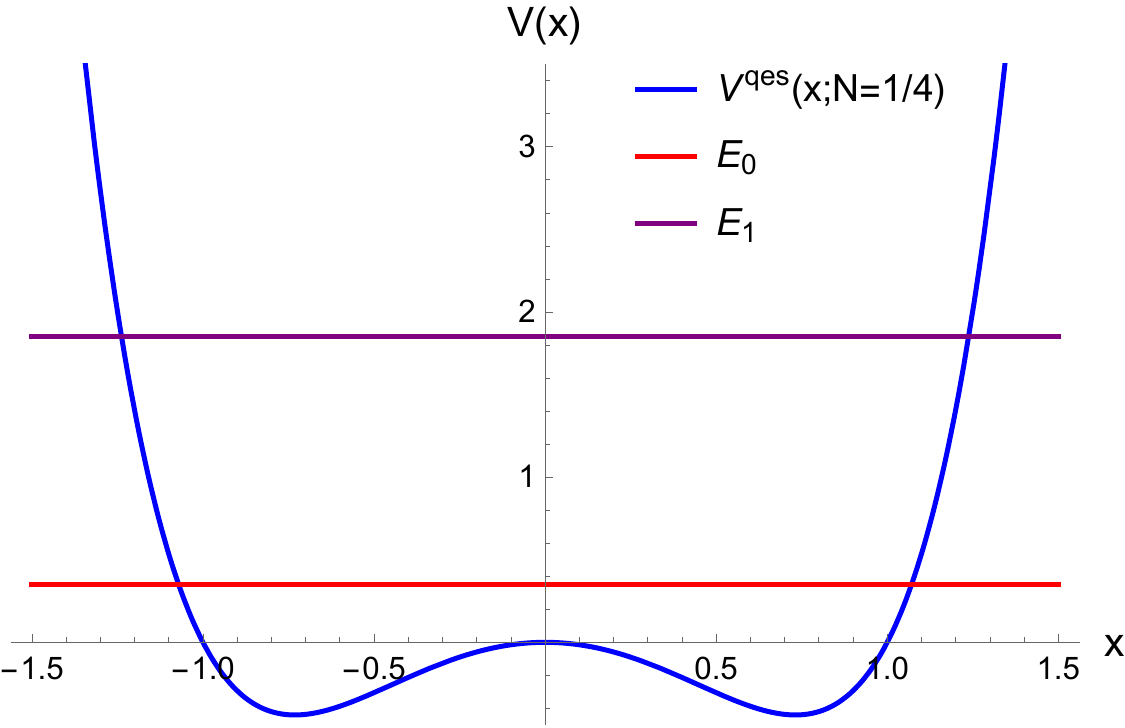}
\caption{Cases $N=0$ (left) and $N=\frac{1}{4}$ (right). The QES sextic potential $V^{\rm(qes)}=\frac{1}{2}\,\big(x^{6} + 2\, x^{4} -  2\,(2N+1) \, x^{2}\,\big)$ is shown (blue line). The ground state energy (in red) and the first excited level (in purple) are displayed as well. No instanton/tunneling effects occur. For $N=0$, the exact analytical ground state is known only, for other levels the numerical results were calculated using the LMMP. For $N=\frac{1}{4}$, no exact analytical solutions exist.}
\label{VeffA}
\end{figure}

\begin{figure}[h]
\centering
\includegraphics[width=7.3cm]{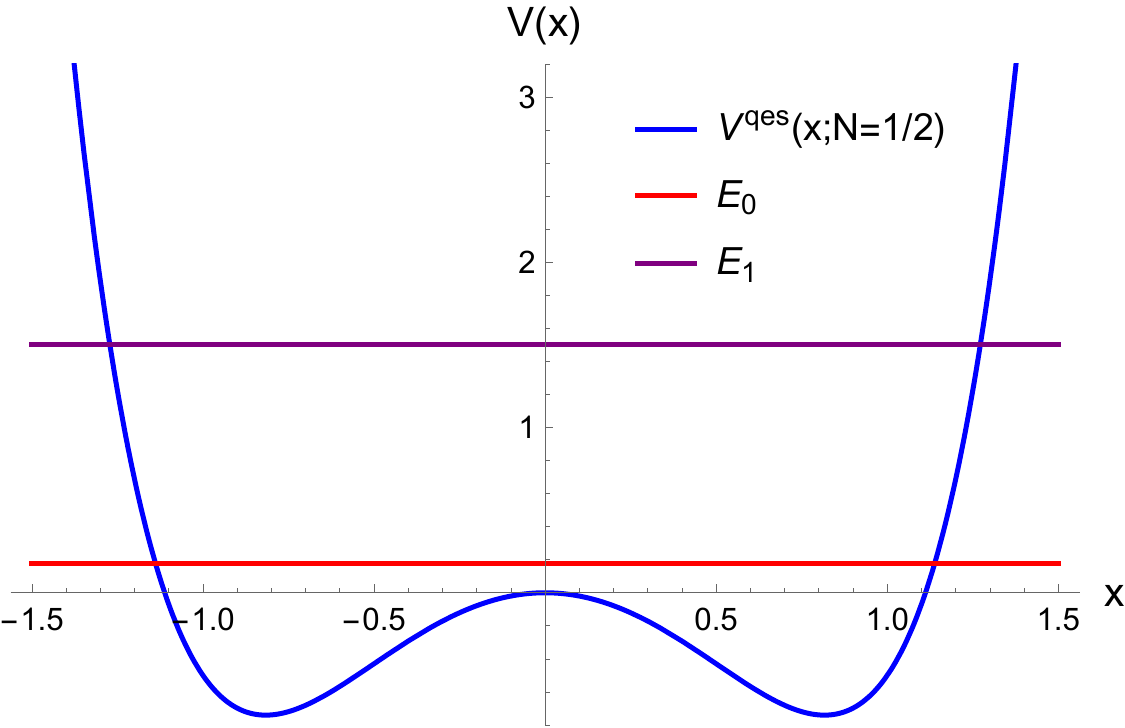} \qquad \includegraphics[width=7.3cm]{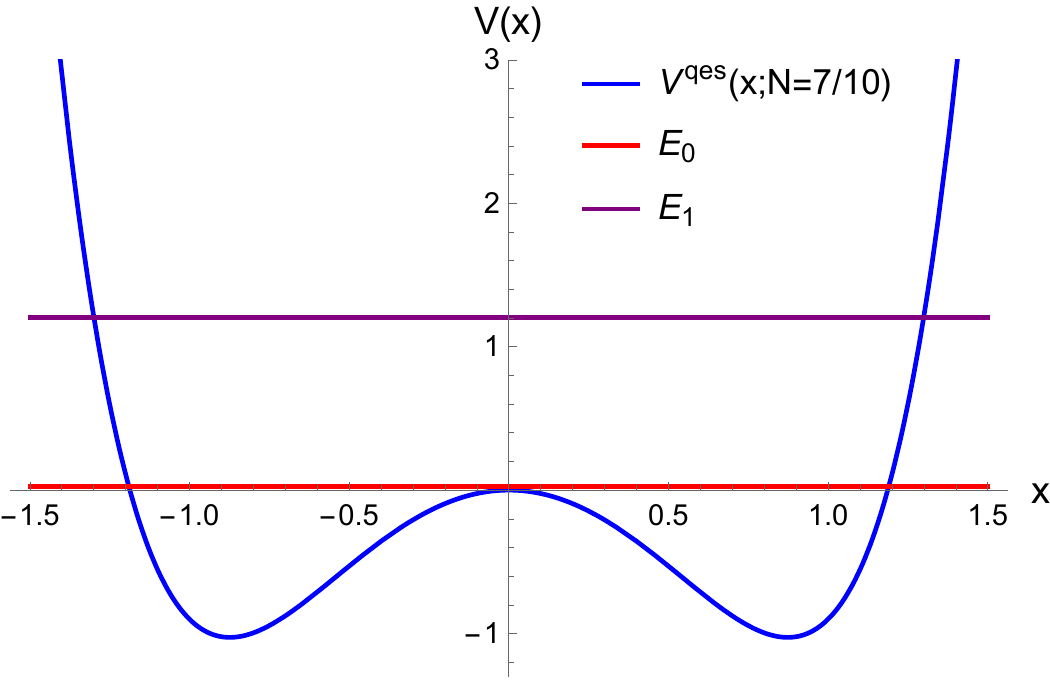}
\caption{Cases $N=\frac{1}{2}$ (left) and $N=\frac{7}{10}$ (right). The QES sextic potential $V^{\rm(qes)}=\frac{1}{2}\,\big(x^{6} + 2\, x^{4} -  2\,(2N+1) \, x^{2}\,\big)$ (blue line). The ground state energy (in red) and the first excited level (in purple) are displayed. No instanton/tunneling effects occur. For $N=\frac{1}{2}$, the exact analytical first excited state is known only, for other levels the numerical results were calculated using the LMMP. For $N=\frac{7}{10}$, no exact analytical solutions exist.}
\label{VeffB}
\end{figure}

For $N<N_{\rm critical} \approx 0.73295$, the ground state energy $E_0=E_0(N)$ is always grater than the maximum of the potential $V^{\rm(qes)}$, thus, no instantons (tunneling) effects would be present, see Figs. \ref{VeffA}, \ref{VeffB}. The instantons effects are beyond the scope of this consideration. 

\begin{figure}[h]
\centering
\includegraphics[width=7.5cm]{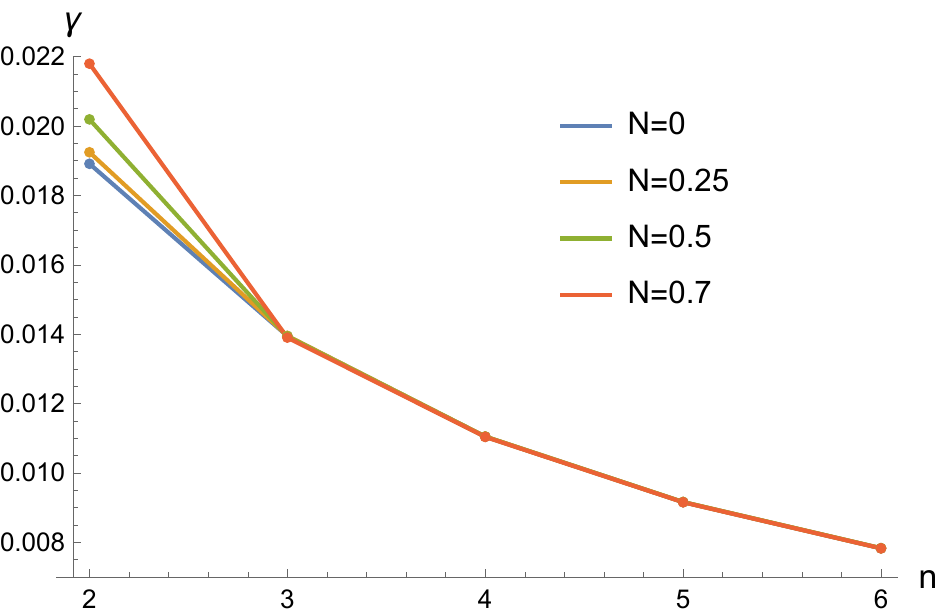} \includegraphics[width=7.5cm]{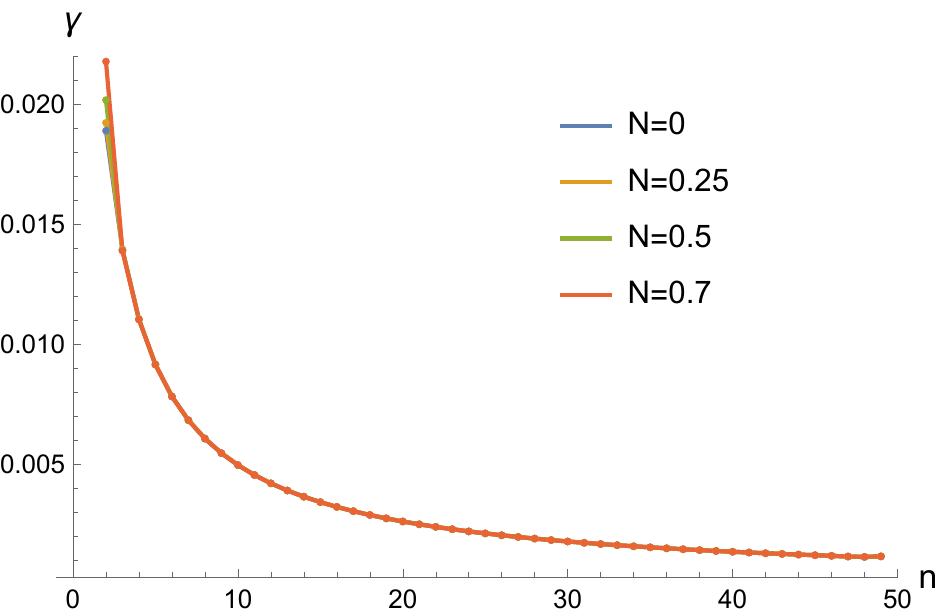} 
\caption{Behaviour of the WKB correction $\gamma=\gamma(n)$ at different values of $N$. Already for $n>2$, the function $\gamma=\gamma(n)$ remains almost unaffected as the parameter $N$ increases from $0$ up to $\frac{7}{10}$.}
\label{gamman}
\end{figure}

\begin{figure}[h]
\centering
\includegraphics[width=10cm]{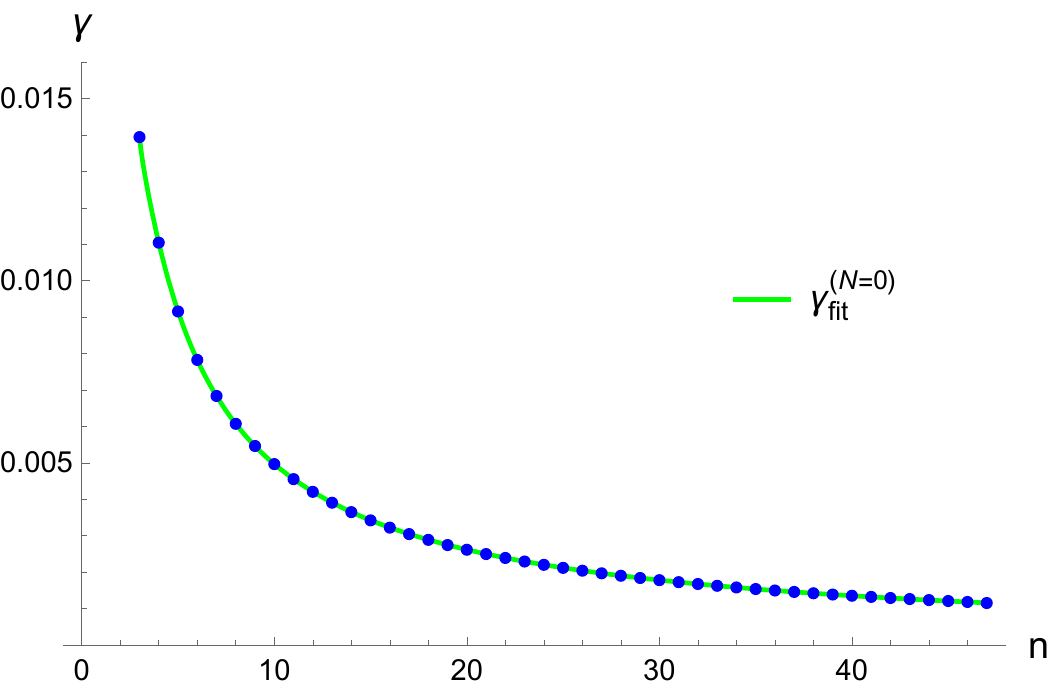}
\caption{Case $N=0$. The analytical interpolation (\ref{interp}) with parameters (\ref{Par0}) corresponds to the green line whilst the bullets (in blue) mark the numerical values $\gamma_{\rm exact}$ of the WKB correction $\gamma=\gamma(n)$.}
\label{IntN0}
\end{figure}

\begin{figure}[h]
\centering
\includegraphics[width=12cm]{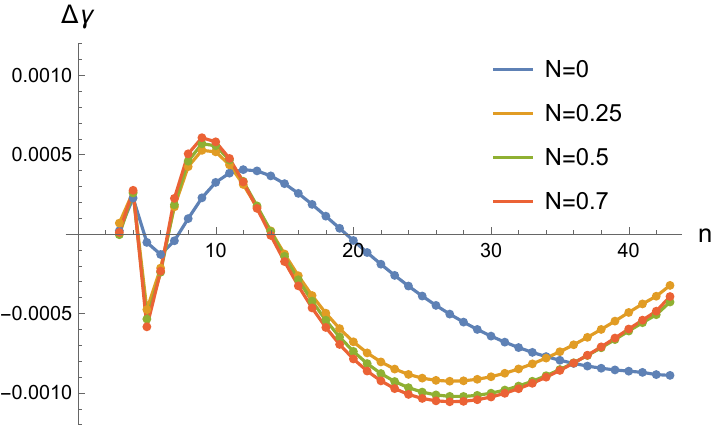}
\caption{Relative error $\Delta\gamma \equiv \frac{\gamma_{\rm exact}-\gamma_{\rm fit}}{\gamma_{\rm exact}}$ \textit{vs} $n$ at different values of $N$. The function $\gamma_{\rm fit}$ is taken as in (\ref{interp}), the values of the interpolation parameters are presented in (\ref{Par0}) and Table \ref{Tabpargammas}, respectively.}
\label{delgam}
\end{figure}

In the cases $N=0,~\frac{1}{4},~\frac{1}{2},~\frac{7}{10},$ we calculate numerically the WKB correction $\gamma=\gamma(N,n)$ for the first 50 states, $n=0,1,2,\,\ldots,50$. The results are displayed in Fig. \ref{gamman}. At fixed $N$, the parameter $\gamma \sim 10^{-2}$ turns out to be a slowly decreasing, smooth function as $n$ increases. Based on \cite{JCRBS} and numerical experiments, it can be fitted by
\begin{equation}
\label{interp}
 \gamma_{\rm fit}\ = \  \frac{a_0 \ + \ a_1\,\tilde n}{ \sqrt{1\,+\,b_1^2\,\tilde n +\,b_2^2\,\tilde n^2 +\,b_3^2\,\tilde n^3+\,b_4^2\,\tilde n^4   }} \ ,
\end{equation}
here $\tilde n\equiv n-2>0$, $a_i=a_i(N)$ and $b_i=b_i(N)$ are interpolating parameters. In the particular case $N=0$, we have
\begin{eqnarray}
\label{Par0}
& a_0 \ \approx \ 0.019202\ , \qquad  \qquad a_1 \ \approx \ 0.0038722\ , \nonumber
\\ &
b_1 \  \approx \ 1.09402 \ , \quad b_2 \ \approx \ 0.672026 \ , \quad b_3 \ \approx \ 0.291328 \ , \quad b_4 \  \approx \ 0.068666 \ ,  
\end{eqnarray}
(see Fig. \ref{IntN0}). For the cases $N=\frac{1}{4},\ \frac{1}{2},\ \frac{7}{10}$, in Table \ref{Tabpargammas} we report the corresponding values of the interpolating parameters in $\gamma_{\rm fit}$ (\ref{interp}), respectively. The fit (\ref{interp}), in comparison with the exact numerical values $\gamma_{\rm exact}$ obtained from (\ref{BSCon}), exhibits a small relative error $|\Delta \gamma| \lesssim 10^{-3}$ in the domain $2 < n <50$ for all values of the parameter $N$ considered, see Fig. \ref{delgam}.

\begin{table}[h]
\centering
\caption{The parameters of the analytical interpolation $\gamma_{\rm fit}$ (\ref{interp}) for the cases $N=\frac{1}{4},~\frac{1}{2},~\frac{7}{10}$.}
\label{Tabpargammas}
\begin{tabular}{c| c |c |c}
\hline
\hline
\hspace{0.2cm}  {\rm parameter}  \hspace{0.4cm}&  $N=\frac{1}4$ & \hspace{0.2cm} $N=\frac{1}{2}$ \hspace{0.2cm} &\hspace{0.1cm } $N=\frac{7}{10}$
\\ \hline
\vspace{-0.2cm} & & & \ \\
$a_0$     & \hspace{0.2cm }  $0.0332512$  \hspace{0.2cm } & $0.0469388$ &  $0.041918$   \   \\ 
\vspace{-0.2cm} & & & \ \\
$a_1$     & $0.0419986$ & $0.0656$ &  $0.0622568$ \  \\
\vspace{-0.2cm} & & & \ \\
$b_1$     & $3.70908$ & $5.57123$ &  $5.08444$ \  \\
\vspace{-0.2cm} & & & \ \\
$b_2$     & $2.96296$  & $4.45283$ &  $4.17168$  \  \\
\vspace{-0.2cm} & & & \ \\
$b_3$     & $2.23779$  & $3.46547$  &  $3.26907$ \  \\
\vspace{-0.2cm} & & & \ \\
$b_4$      & $0.749137$  & $1.17011$ &  $1.11086$ \  \\
\vspace{-0.2cm} & & & \ \\
\hline
\hline
\end{tabular}
\end{table}

\clearpage

\section{Energies: analytical interpolations}

\vspace{0.2cm}

\begin{figure}[h]
\centering
\includegraphics[width=7.5cm]{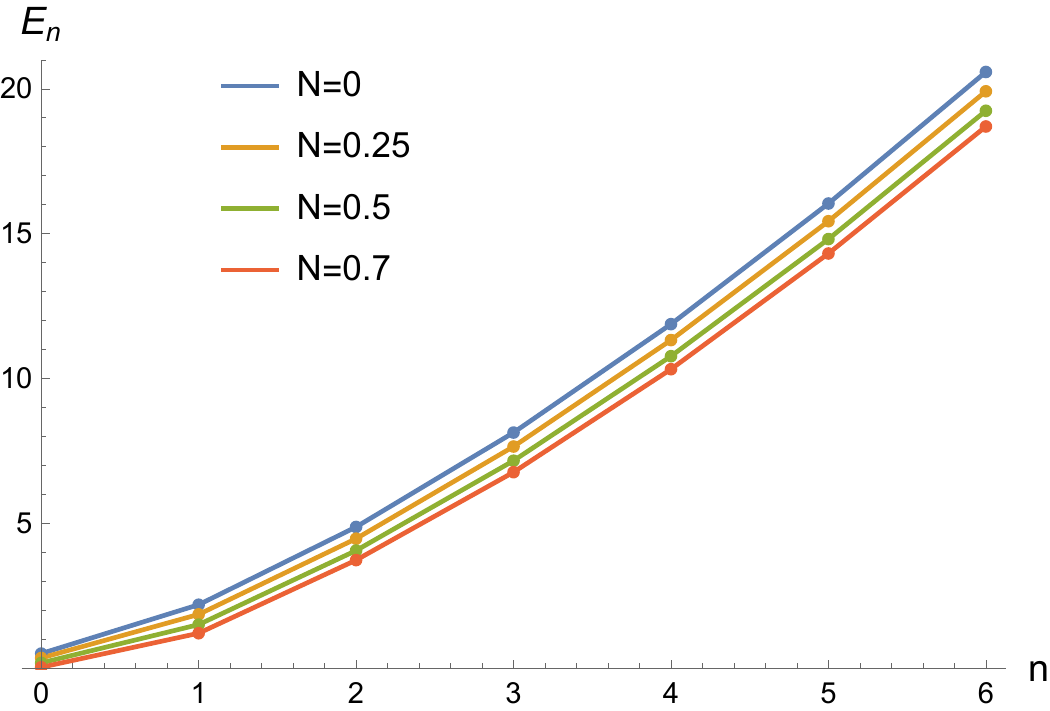} \includegraphics[width=7.5cm]{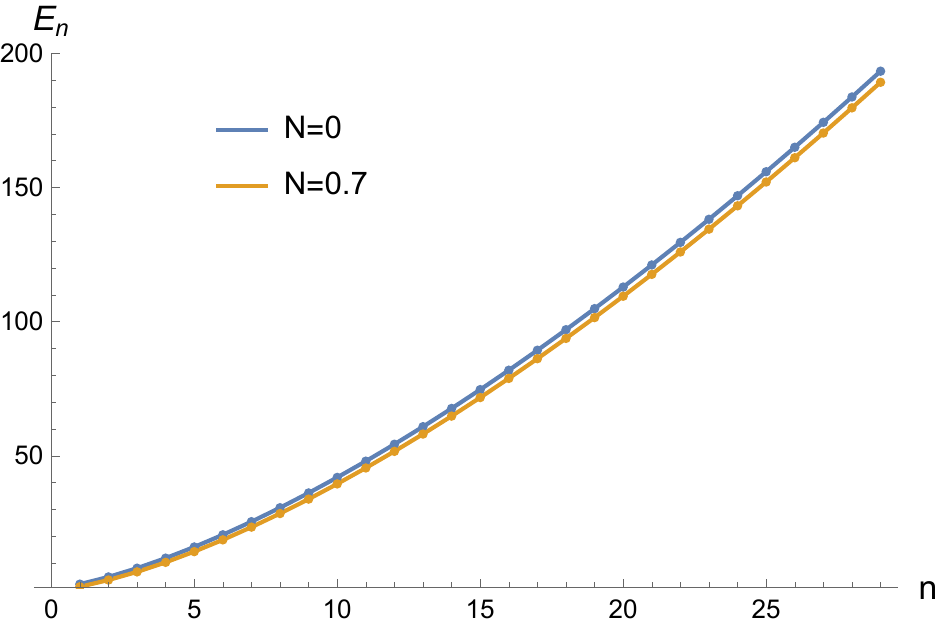}
\caption{The energy $E_n$ \textit{vs} $n$ for the cases $N=0,\frac{1}{4},\frac{1}{2},\frac{7}{10}$. Results are displayed in atomic units.}
\label{Ener}
\end{figure}

As mentioned in Section \ref{sec WKB}, for the QES sextic potential (\ref{Vred}) using the LMMP we computed at $N=0,\frac{1}{4},\frac{1}{2},\frac{7}{10}$ the energy $E_n$ for the first 50 states. At fixed $N$, it is a slowly increasing, smooth function as $n$ grows, see Fig. \ref{Ener}. In the domain $n \in [0,50]$ the energy varies from $\sim 1$ up to $\sim 410$. At fixed $n$, the energy decreases monotonously as the parameter $N$ changes from $0$ up to $N=0.7$. 

\begin{figure}[h]
\centering
\includegraphics[width=8cm]{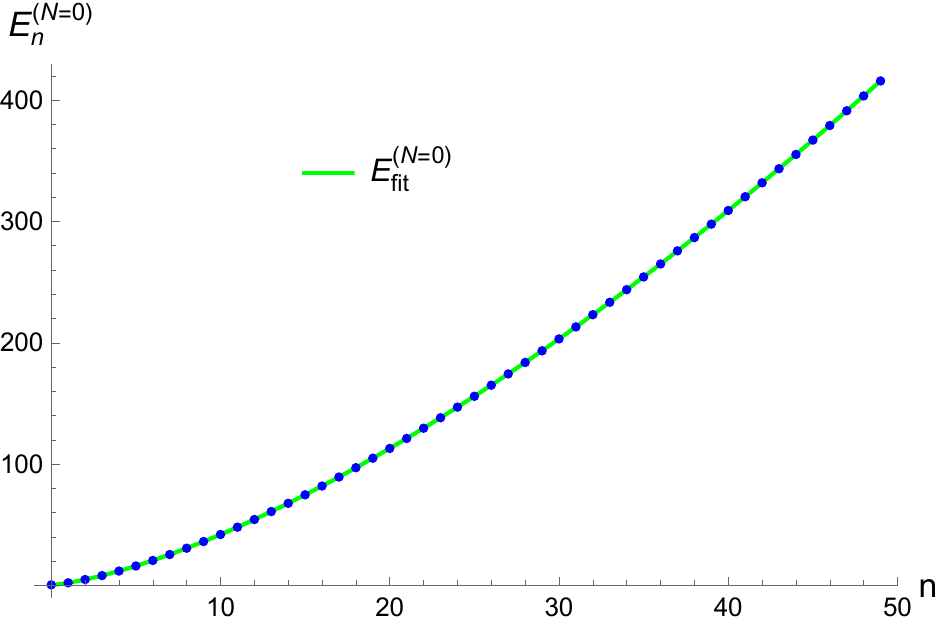}
\caption{Case $N=0$. The analytical interpolation (\ref{Einter}) with parameters (\ref{Par00}) corresponds to the green line whilst the bullets (in blue) mark the numerical values obtained using the LMMP.}
\label{E0N0a}
\end{figure}

At fixed $N$, the energy $E_n=E(n)$ can be interpolated by
\begin{equation}
E_{\rm fit}(n)\ = \  E_0\,\tilde n \ + \ \sqrt{\tilde n-1}\,\frac{(a_0 \,+\,A_1\,\tilde n\,+\,A_2\,\tilde n^2\,+\,A_3\,\tilde n^3\,+\,A_4\,\tilde n^4\,+\,A_5\,\tilde n^5\,+\,A_6\,\tilde n^6)}{1\,+\,B_1^2\,\tilde n\,+\,B_2^2\,\tilde n^2\,+\,B_3^2\,\tilde n^3\,+\,B_4^2\,\tilde n^4\,+\,B_5^2\,\tilde n^5}\ ,
\label{Einter}
\end{equation}
where $\tilde n \equiv n+1>0$, $A_i=A_i(N)$ and $B_i=B_i(N)$ are interpolating parameters. At $n=0$ ($\tilde n=1$) the analytical interpolation $E_{\rm fit}$ reproduces the exact numerical value of the ground state energy $E_{\rm fit} \rightarrow E_0$ whereas at large $n \rightarrow \infty$ (thus, $\tilde n \rightarrow \infty$) we reproduce the asymptotic behaviour $E \propto n^{3/2}$. If $n \rightarrow \infty$, in the semi-classical limit we consider large distances $x \rightarrow \infty$ where the sextic potential behaves as $V^{\rm (qes)} \sim \frac{x^6}{2}$. Standard Bohr-Sommerfeld quantization condition provides the dominant term 
\[
E_n \ \sim \ \frac{1}{2} \pi ^{3/4} \left(\frac{\Gamma \left(\frac{5}{3}\right)}{\Gamma \left(\frac{7}{6}\right)}\right)^{3/2}\, n^{3/2}  \ = \ 1.13254\, n^{3/2}   \ ,
\]
in the limit $n \rightarrow \infty$. Specifically, for the interpolation (\ref{Einter}) with $N=0$ we obtain

\begin{align}
\label{Par00}
    &  A_0 \ = \ -303.678, \quad \ A_1 \ = \ -13.8988, \quad  \ A_2 \ = \ 185.841,  \quad  \ A_3 \ = \ 22.0053, \quad  \ A_4 \ = \  13.4821,  \nonumber \\ 
    &   A_5 \ = \  0.777246, \quad   \ \ A_6 \ = \  1.06539, \nonumber \\ 
    & B_1 \ = \ 16.0303, \quad  \ \ \ B_2 \ = \ 4.22639, \quad  \ B_3 \ = \ 3.82635, \quad  \ B_4 \ = \ 1.17858, \quad  \ B_5 \ = \ 0.969174, 
\end{align}
see Fig. \ref{E0N0a}. In this case, for the asymptotic behaviour at $n \rightarrow \infty$ we found that $E_{\rm fit} \sim 1.13424\, n^{3/2}$.

\begin{figure}[h]
\centering
\includegraphics[width=7.8cm]{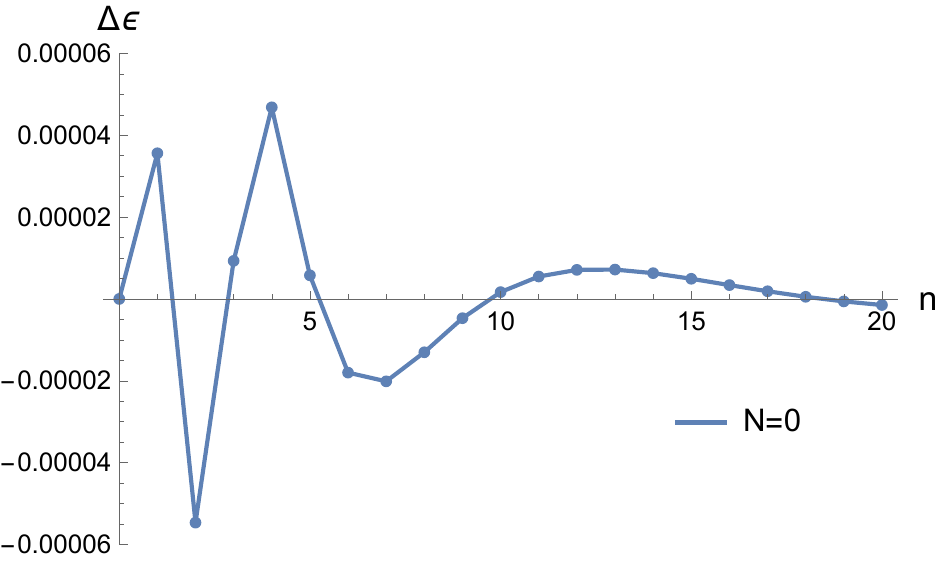} \includegraphics[width=7.8cm]{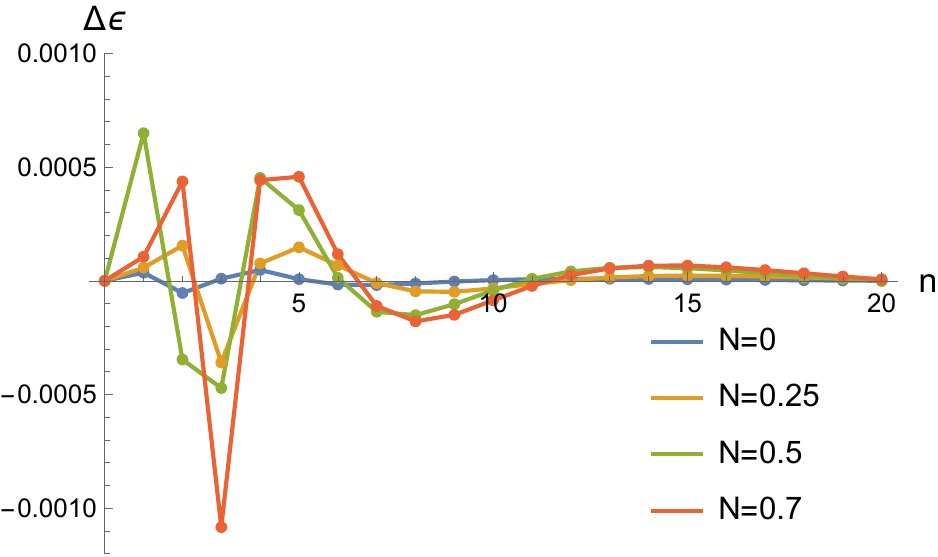}
\caption{Relative error $\Delta \epsilon \equiv \frac{E_{\rm exact}-E_{\rm fit}}{E_{\rm exact}}$ \textit{vs} $n$. The function $E_{\rm fit}$ is taken as in (\ref{Einter}) whilst the values of the interpolation parameters are displayed in (\ref{Par00}) and Table \ref{Tab2}.}
\label{delep}
\end{figure}

In Table \ref{Tab2} the corresponding values of the interpolating parameters in $E_{\rm fit}$ (\ref{Einter}) are shown as a function of $N$. The fit (\ref{Einter}), in comparison with the exact numerical values $E_{\rm exact}$ obtained using the LMMP, exhibits a small relative error $|\Delta \epsilon| \lesssim 10^{-3}$ in the whole domain $2 < n <50$ for all values of the parameter $N$ considered, see Fig. \ref{delep}. In the particular case $N=0$, this error reduces up to $|\Delta \epsilon |\lesssim 10^{-5}$. 

\clearpage

\begin{table}[h]
\centering
\caption{The parameters of the analytical interpolation $E_{\rm fit}$ (\ref{Einter}) for the cases $N=\frac{1}{4},\,\frac{1}{2},\,\frac{7}{10}$. The ratio $A_6/B^2_5$ defines the asymptotic behaviour at $n \rightarrow \infty$ in $E_{\rm fit}$ (see text).}
\label{}
\begin{tabular}{c|  c |c |c}
\hline
\hline
\hspace{0.2cm}  {\rm parameter}  \hspace{0.4cm}& $N=\frac{1}4$ & \hspace{0.2cm} $N=\frac{1}{2}$ \hspace{0.2cm} &\hspace{0.1cm } $N=\frac{7}{10}$\\ \hline
$A_0$     & $-2961.78$ \hspace{0.2cm } & $-4024.34$   &  $-6563.79$ \   \\
$A_1$     & $407.572$ & $1776.87$ & $3617.77$ \  \\
$A_2$     & $687.568$ & $553.412$ & $25.2448$  \  \\
$A_3$     & $684.966$ & $210.095$ & $443.209$  \   \\
$A_4$     & $-1.90414$  & $13.202$ & $20.7142$  \ \\
$A_5$     & $0.755926$  & $1.7131$ & $2.55102$  \  \\
$A_6$     & $1.06975$  & $1.07253$ &  $1.07404$ \  \\
$B_1$     & $40.1137$  & $37.9358$ & $40.0919$  \  \\
$B_2$     & $22.2057$  & $5.64541$  & $11.9183$  \  \\
$B_3$      & $3.02188$  & $5.36947$ & $6.51999$ \  \\
$B_4$      & $0.936421$  & $1.09358$ & $1.18494$   \ \\
$B_5$      & $0.967749$  & $0.965024$ &  $0.962401$  \ \\
\hline
\hline
$A_6/B^2_5$      & $1.14224$  & $1.15169$ &  $1.1596$  \ \\
\hline
\hline
\label{Tab2}
\end{tabular}
\end{table}

\section{ Supersymmetric quantum mechanics and intertwining relations}
\label{susyin}

\vspace{0.2cm}

Supersymmetric quantum mechanics (SUSY) is a technique used to expand the family of solvable potentials in quantum mechanics \cite{David2010}. It relates two different Hamiltonian quantum systems, $H_0$ and $H_1$, by means of an intertwining operator $A^+_1$ as follows: 
\begin{equation} \label{interwining}
    H_1 A^+_1 = A^+_1 H_0, \quad \text{where} \quad H_i = - \frac{1}{2} \frac{d^2}{dx^2} + V_i(x), \quad i=0,1 \ .  
\end{equation}
This intertwining relation allows us to map solutions of the eigenvalue equation $H_0\, \psi = E \,\psi$ to solutions of the isospectral problem $H_1\, \phi = E \,\phi$ (except possibly for certain energy eigenstates). Naturally, the relation (\ref{interwining}) imposes certain restrictions on $H_1$. To see this, let us consider the simplest case when $A^+_1$ is a first-order differential operator
\begin{equation}
\label{Ap1}
    A^+_1 = \frac{1}{\sqrt{2}}\left( -\frac{d}{dx} + W(x) \right) = \frac{1}{\sqrt{2}} \left( -\frac{d}{dx} + (\ln u(x))' \right), 
\end{equation}
where $W(x)$ or, equivalently, $u(x)$ are functions to be found. We denote $u'(x)$ as $\frac{d u(x)}{dx}$. In the literature, $W(x)$ is known as superpotential and $u(x)$ as the seed function. By substituting the explicit form of $A^+_1$ \eqref{Ap1} in \eqref{interwining}, we obtain the conditions 
\begin{eqnarray}
    H_1 = H_0 - \left(\ln u(x) \right)''\,, \quad H_0 \,u(x) = \epsilon\, u(x)\ , \quad u(x) \neq 0, 
\end{eqnarray}
here $\epsilon$ is a constant; i.e., $u(x)$ must be a nodeless solution for the initial Schr\"odinger equation with Hamiltonian $H_0$. Once $u(x)$ is chosen, $H_1$ is fixed.  

\section{SUSY partners of the sextic potentials}

\vspace{0.2cm}

Recently, in the work \cite{Escobar2024}, the present authors studied the algebraic structure of the SUSY partners $V_1(x)$ of the sextic potential $V_0=V^{\rm qes}(x) = \frac{1}{2}(  \nu^{2}\,x^{6} +  2\,\nu\, \mu\, x^{4} +  (\mu^{2}\,-\,(4N+3)\nu) \ x^{2})$ as a function of $N$. Accordingly, the central object was the Hamiltonian $H_0 = -\frac{1}{2}\frac{d^2}{dx^2} \, + \, V^{\rm qes}(x)$.
Here, in the $\mathbb{Z}_2$-invariant variable $z=x^2$, we further elaborate on the relevant intertwining relation between $H_0$ and $H_1$ at the level of the algebraic gauge rotated operators, $h_0$ and $h_1$, which govern the polynomial part of the exact QES eigenfunctions, respectively. Explicitly, using the gauge factor 
\begin{equation}
    \Gamma(z) = \exp \left(-\frac{\nu}{4}z^2 - \frac{\mu}{2}z\right) \ ,
\end{equation}
the gauge-rotated Hamiltonian $h_0=\Gamma^{-1} \,H_0 \,\Gamma$ takes the form: 
\begin{equation}
    h_0\ = \ -2\,z \,\frac{d^2}{dz^2} \ + \ \left( 2 \,\nu\, z^2 + 2\,\mu\, z -1 \right) \frac{d}{dz} \ - \ 2\, N\, \nu\, z \ + \  \frac{\mu}{2}\ .
\end{equation}
If the parameter $N$ takes positive integer values, the spectral problem $h_0 \,P(z)=E\,P(z)$ has $(N+1)$ polynomial eigenfunctions $P_j^{(N)}(z)$, $j=0, 1, 2, N$. Accordingly, the QES solutions of the original Schr\"odinger equation $H_0 \,\psi = E\, \psi$ become $\psi_j(x)= \Gamma(z=x^2) P_j^{(N)}(z=x^2)$. The above operator $h_0$ possesses a hidden $\mathfrak{sl}_2(\mathbb{R})$ algebra.

Now, we can perform a first-order SUSY transformation of $H_0$, described above in section \ref{susyin}, using as a seed function the exact (analytical) ground state $\psi_0(x)= \Gamma(x^2) P_0^{(N)}(x^2)$. That way, we obtain the SUSY partner Hamiltonian $H_1$. For this operator $H_1$, the $N$ exact QES solutions can be written as $\phi_i(x) \ = \ \Gamma_N(x)\,\mathcal{P}_i(x)$, $i=0, 1, 2, (N-1)$ , where the factor $\mathcal{P}_i(x)$
is a polynomial odd-function in $x-$variable whilst $\Gamma_N$ is the non-polynomial part. Further details can be found in \cite{Escobar2024}. 

Similarly to $H_0$, one can introduce the operator $h_1= \Gamma_N^{-1} \,H_1\, \Gamma_N$ which describes the polynomial QES solutions  $\mathcal{P}_i(x)$ of $H_1$. From the intertwining relation \eqref{interwining}, we can immediately derive the SUSY partner of $h_0$, namely
\begin{equation} \label{pol intertwining}
    h_1 \,\mathcal{A}_1^+ \ = \  \mathcal{A}_1^+\, h_0 \ , 
\end{equation}
with intertwining operator  $\mathcal{A}_1^+ = \Gamma_N^{-1} A_1^+ \,\Gamma$ and $A_1^+$ defined in (\ref{Ap1}). 

Eventually, from the equation $h_0\, P(z)=E\,P(z)$ and \eqref{pol intertwining} we obtain that $\mathcal{P}(\sqrt{z})= \mathcal{A}_1^+ P(z)$ solves  $h_1 \mathcal{P}(\sqrt{z})=E\, \mathcal{P}(\sqrt{z})$.
In variable $z$, the operators $\mathcal{A}_1^+$ and $h_1$ read
\begin{eqnarray}
    \mathcal{A}_1^+ &=& -\sqrt{2\, z}\, p(z) \,\left( \frac{d}{dz} \ + \ 2 \,\frac{\dot p(z)}{p(z)} \right), \nonumber \\
    h_1 &=& h_0 \ + \  4\,z\, \frac{\dot p(z)}{p(z)} \frac{d}{dz} \ - \ 2\,z \frac{\ddot p(z)}{p(z)} - (2 \,\nu\, z^2 + 2 \,\mu\, z+1) \frac{\dot p(z)}{ p(z)} \ + \  3\, \nu\, z \ + \  \mu \ , 
\end{eqnarray}
where we have used the notation $p(z)=P_0^{(N)}(z)$ and $\dot p (z) = \frac{d p(z)}{dz}$. Except for the case $N=0$, no Lie algebraic structure is known in $h_1$.

\section{SUSY partners of the quasi-exactly solvable Morse potential}

\vspace{0.2cm}

In this Section, we consider the QES Morse potential \cite{TURBINER20161}
\begin{equation}
 V_0(x;N) \ = \   \frac{1}{2}\big( a^2\, e^{-2 \,\alpha \, x} \ - \ a\, e^{-\alpha\,  x} (2 \,b+\alpha \, (2\,N+1)\,)  \ + \ (N \alpha +b)^2\big) \ ,
\end{equation}
a QES system of the first type, $x \in (-\infty,\infty)$ where  $a>0,\,b,\,\alpha>0$ and $N$ are parameters. Let us take the \textit{fermionic} Hamiltonian

\begin{equation}
    H_0 \ = \ -\frac{1}{2}\frac{d^2}{dx^2} \ + \ V_0(x;N)  \ .
\label{HMor}    
\end{equation}
In the variable
\begin{equation}
    z \ = \ e^{-\alpha\,x} \ ,
\end{equation}
the operator (\ref{HMor}) can be transformed to a quantum ``top''  \cite{TURBINER20161}; namely, it can be rewritten as a constant coefficient quadratic combination in the first order differential operators
\begin{equation}
 {\cal J}^+(N) \ \equiv \ z^2\,\partial_z \ - \ N\,z \ , \qquad {\cal J}^0(N) \ \equiv \ z\,\partial_z \ - \ 
 \frac{N}{2} \ , \qquad {\cal J}^- \ \equiv \ \partial_z \ , 
\end{equation}
which span, for any value of the parameter $N$, the $\mathfrak{sl}_2(\mathbb{R})$ algebra. Explicitly, using the gauge factor
\begin{equation}
    \Gamma(z) \ = \ e^{-\frac{a}{\alpha }\,e^{-\alpha  x}-b\, x}
    \ = \   e^{-\frac{a}{\alpha }\,z+\frac{b}{\alpha}\,\log(z)}  \ ,
\end{equation}
one obtains the gauge-rotated algebraic operator 
\begin{eqnarray}
 h^{(N)}_0\ &\equiv & \ {\Gamma}^{-1}\,H_0\,\Gamma  \nonumber \\ 
&  = &\ -\frac{1}{2} \left(2 \,a \,\alpha \, N z\,+\,b^2-2 c\right) \  - \ \frac{1}{2} \alpha  \,z\,(-2 \,a z+\alpha +2\, b)\,\partial_z   \ - \ \frac{1}{2} \,\alpha ^2 \,z^2\, \partial^2_z
\ ,
\label{oph}
\end{eqnarray}
or, equivalently, in Lie algebraic form
\begin{equation}
h^{(N)}_0 \ = \ -\frac{\alpha ^2}{2}\,{\cal J}^+\,{\cal J}^- \ + \ a\,\alpha\,{\cal J}^+ \ - \ \frac{1}{2} \,\alpha\,  (2 b+\alpha \,(N\,+\,1)\,)\,{\cal J}^0 \ + \ \frac{1}{2} \left(2\, c-b^2\right) \ .
\label{ophLie}
\end{equation}
At fixed $N$, the spectral problem 
\begin{equation}
h^{(N)}_0\,P(z) \ = \ \,E\,P(z) \ ,
\end{equation}
admits $(N+1)$ polynomial solutions $P_i(z)$ in the $z-$variable. Accordingly, the original Hamiltonian operator $H_0$ (\ref{HMor}) possesses $(N+1)$ eigenfunctions in the form 
\begin{equation}
\psi_i \ = \  \Gamma(z)\,P_i(z) \ , \quad i=1,2,\ldots,N \ . 
\end{equation}
In $z-$variable, the original Hamiltonian operator $H_0$ (\ref{HMor}) reads
\begin{equation}
    H_0 \ = \ -\frac{1}{2}\alpha^2\,(z^2 \,\partial^2_z \ + \ z\,\partial_z) \ + \   \frac{1}{2} \left(a^2 \,z^2 \ - \ a \,(\alpha\,(2N+1) 
 \, + \,2 b)\, z+2\,c\right) \ .
\end{equation}
Below, we present concrete examples.

\subsection{Case $N=0$}

\vspace{0.2cm}

\begin{figure}[t]
\centering
\includegraphics[width=7cm]{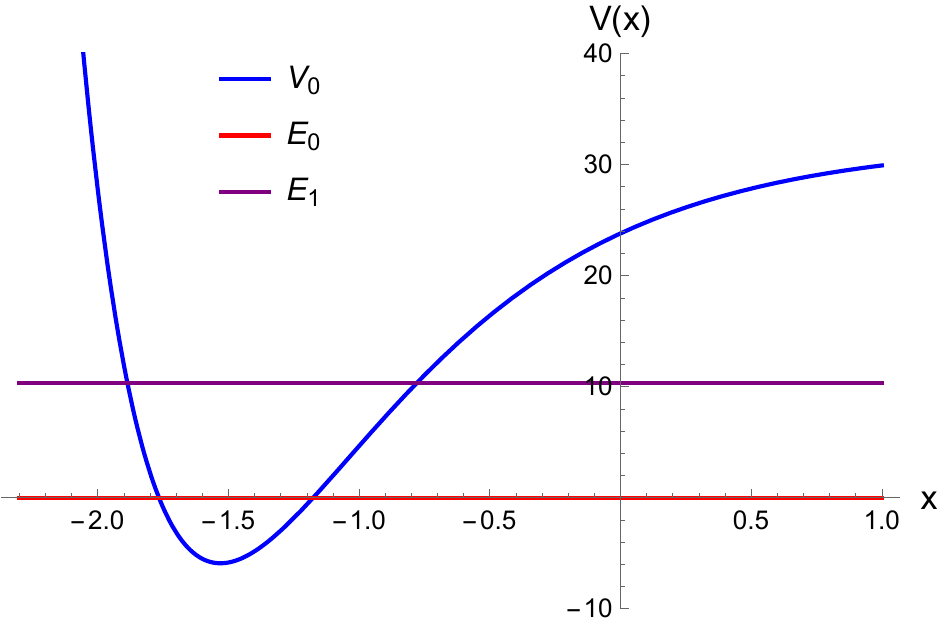}
\caption{$a=1, b=8,\alpha=\sqrt{2}$ and $N=0$, the potential $V_0$ (blue curve) in (\ref{HMor}) and the first two energy levels $E_0$ and $E_1$.}
\label{F9}
\end{figure}

For $N=0$, the potential $V_0$ in (\ref{HMor}) reads
\begin{equation}
 V_0(x;N=0) \ = \   \frac{1}{2} \left(a^2 e^{-2 \,\alpha\,  x} \ - \ a \,(\alpha 
 \, + \,2 b)\, e^{-\alpha  \,x}+b^2\right) \ = \   \frac{1}{2} \left(a^2 \,z^2 \ - \ a \,(\alpha 
 \, + \,2 b)\, z+b^2\right) \ .
\end{equation}
The ground state function of (\ref{HMor}) is given by
\begin{equation}
    \psi_0^{(N=0)} \ = \ \Gamma  \ = \ e^{-\frac{a}{\alpha }\,e^{-\alpha  x}-b\, x} \ = \   e^{-\frac{a}{\alpha }\,z+\frac{b}{\alpha}\,\log(z)}\ ,
\end{equation}
$P(z)=1$, with zero energy
\begin{equation}
    E_0^{(N=0)} \ = \ 0 \ .
\end{equation}
In this case,
\begin{equation}
h^{(N=0)}\ \equiv \ {\big(\psi_0^{(N=0)}\big)}^{-1}\,H_0\,\psi_0^{(N=0)} \ = \ \frac{1}{2}\,\alpha\,\bigg(\,-\alpha\,{\cal J}^+\, {\cal J}^- \ + \ 2\,a\,{\cal J}^+ \ - \ (\alpha+2\,b)\,{\cal J}^0 \bigg) \ . 
\end{equation}
For instance, taking $a=1,~ b=8,~\alpha=\sqrt{2}$, the next lowest energies levels are $E_1=10.313708498985$, $E_2=18.62741699797$, $E_3=24.94112549695$, $E_4=29.25483399594$, $E_5=31.56854249492$, see Fig. \ref{F9}. Interestingly, for the QES Morse potential, it can be checked that the semiclassical WKB method reproduces the exact quantum mechanical results. Thus, the WKB correction vanishes $\gamma =0$ for all the bound states. Explicitly, 
\begin{equation}
\begin{aligned}
\label{}
\int_{x_1(E)}^{x_2(E)}\sqrt{2\,\big(\,E \ - \ V_0(x,N=0)\,\big)\,dx} & \ = \ \frac{\pi  \left(\alpha -2 \sqrt{b^2-2 \,E}+2 b\right)}{2 \,\alpha }
\\ & \ =
\pi\,(n+1/2)\ ,
\end{aligned}
\end{equation}
($\gamma=0$) which leads to the exact quantum spectra $E_n=-\frac{1}{2} \alpha\,  n \,(\alpha \, n-2\, b)$.
For the treatment of the exactly solvable case $V_{ ES} \propto (e^{-2\,\alpha\,x}\,-\,2\,e^{-\alpha\,x}\,)$, see \cite{Ivanov1997}.

The associated SUSY partner potential can be derived immediately
\begin{equation}
    V_1(x;N=0) \ = \ \frac{1}{2} \,a^2\, e^{-2 \alpha  x}\ + \ a\,e^{-\alpha \, x} \left(\frac{ \alpha }{2}- b\right) \ + \ \frac{b^2}{2} \ = \ \frac{1}{2} (\,a^2\,z^2\ + \ a\,z\, \left(\alpha -2\, b\right) \ + \ b^2\,)\ .
\label{PV1}    
\end{equation}
Also, at $N=0$ the \textit{bosonic} Hamiltonian 
\begin{equation}
    H_1^{(N=0)} \ = \ -\frac{1}{2}\frac{d^2}{dx^2} \ + \ V_1(x;N=0)  
 \ =  \ -\frac{1}{2}\frac{d^2}{dx^2} \ + \ V_0(x,N=-1) \ ,
\label{}    
\end{equation}
can be transformed into a quantum ``top'' of the form (\ref{ophLie}) but with a negative integer (cohomology) parameter $N=-1$. Therefore, similar to the QES sextic potential \cite{Escobar2024} the SUSY partner Hamiltonian $H_1^{(N=0)}$ with potential (\ref{PV1}) still possesses
a hidden $\mathfrak{sl}_2(\mathbb{R})$ Lie algebraic structure.

\begin{figure}[t]
\centering
\includegraphics[width=7cm]{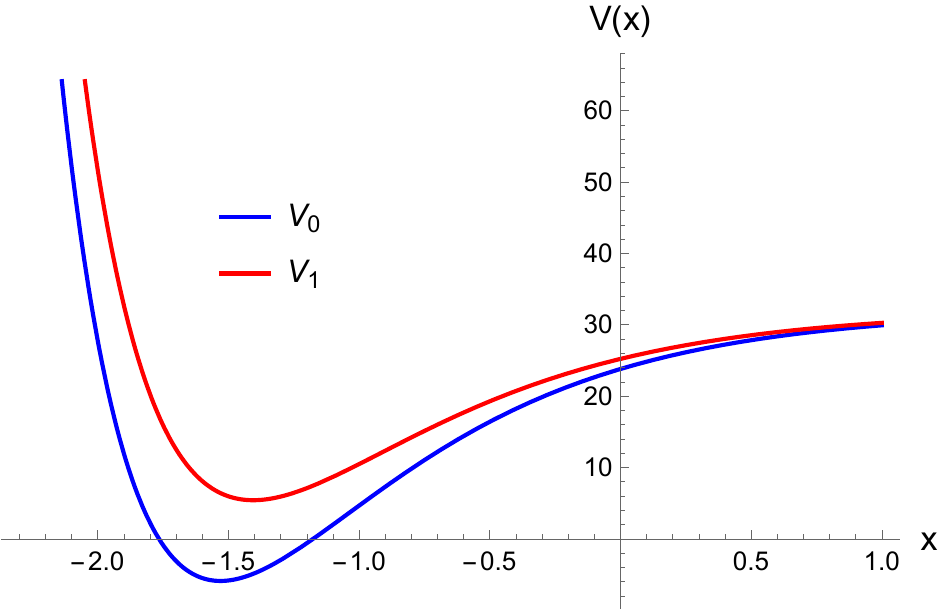}
\caption{For fixed $a=1, b=8,\alpha=\sqrt{2}$ and $N=0$, the confining potential $V_0$ (blue curve) in (\ref{HMor}) and its susy partner $V_1(x)$.}
\label{F10}
\end{figure}

\subsection{Case $N=1$}

\vspace{0.2cm}

For $N=1$, we have
\begin{equation}
    V_0 \ = \ \frac{1}{2} \left(a^2 e^{-2 \alpha  x}-a (3 \alpha +2 b) e^{-\alpha \,x}+ (\alpha +b)^2\,\right) \ = \ \frac{1}{2} \left(a^2 \,z^2-a\,z\, (3 \alpha +2 b)+ (\alpha +b)^2\,\right) \ ,
\end{equation}
see Fig. \ref{F10}. The exact solutions are the ground state function given by 
\begin{equation}
    \psi_0^{(N=1)} \ = \ e^{-\frac{a}{\alpha }\,e^{-\alpha  x}-b\,x }\,e^{-\alpha \, x} \ = \   e^{-\frac{a}{\alpha }\,z+\frac{b}{\alpha}\,\log(z)}\,z \ ,
\end{equation}
$P(z)=z$, again with zero energy
\begin{equation}
    E_0^{(N=1)} \ = \ 0 \ ,
\end{equation}
and the first excited state
\begin{equation}
    \psi_1^{(N=1)} \ = \ e^{-\frac{a}{\alpha }\,e^{-\alpha  x}-b\,x }(e^{-\alpha \, x} \ - \ \frac{\alpha +2 \,b}{2\, a} )\ = \  e^{-\frac{a}{\alpha }\,z+\frac{b}{\alpha}\,\log(z)}\,(z \ - \ \frac{\alpha +2 \,b}{2\, a} )\ ,
\end{equation}
thus, $P(z)=z -\frac{\alpha +2 \,b}{2\, a}$, with energy
\begin{equation}
    E_1^{(N=1)} \ = \ \frac{1}{2}\, \alpha\,  (\alpha +2 b) \ .
\end{equation}
Eventually, the associated susy partner potential satisfies that
\begin{equation}
    V_1(x;N=1) \ =  \ V_0(x,N=0)\ .
\label{}    
\end{equation}
 
\begin{figure}[t]
\centering
\includegraphics[width=7cm]{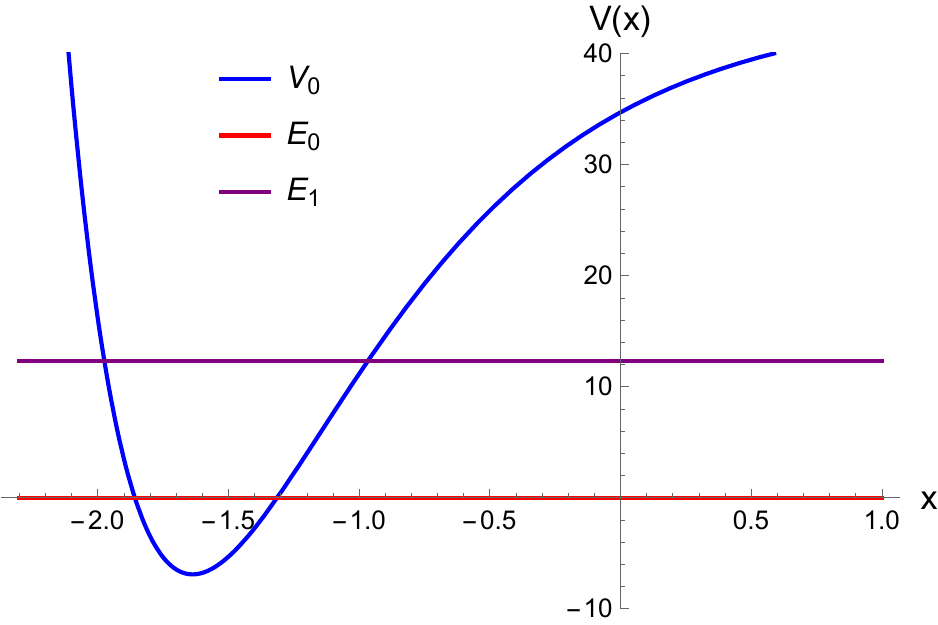}
\caption{$a=1, b=8,\alpha=\sqrt{2}$ and $N=1$, the potential $V_0$ (blue curve) in (\ref{HMor}) and the first two energy levels $E_0$ and $E_1$.}
\end{figure}

\begin{figure}[t]
\centering
\includegraphics[width=7cm]{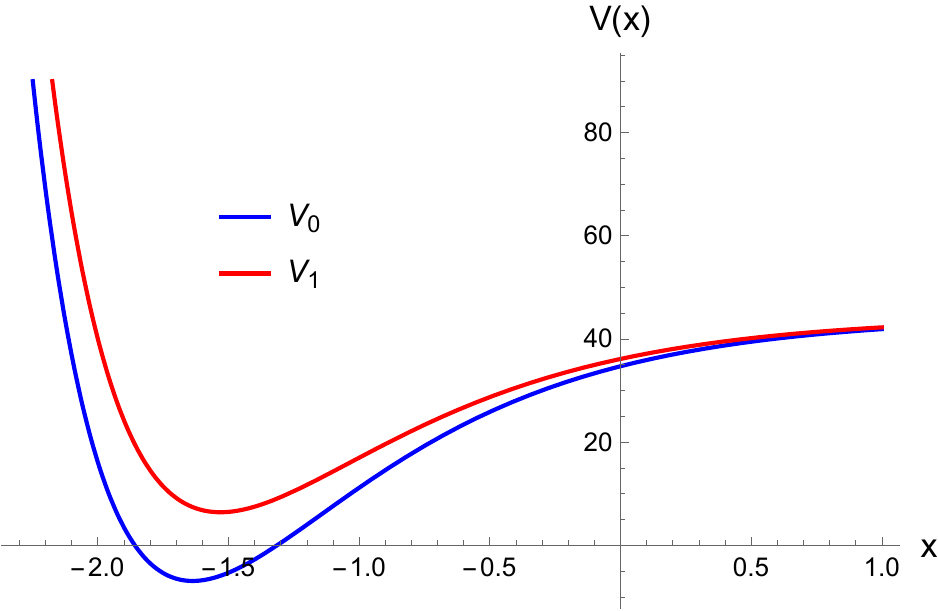}
\caption{For fixed $a=1,\, b=8,\,\alpha=\sqrt{2}$ and $N=1$, the confining potential $V_0$ (blue curve) in (\ref{HMor}) and its susy partner $V_1(x)$.}
\end{figure}

\subsection{Case arbitrary integer $N$}

\vspace{0.2cm}

For arbitrary $N$, we have
\begin{equation}
\begin{aligned}
V_0 \ & =  \ \frac{1}{2} \left(a^2 e^{-2 \alpha  x}-a \big(\alpha(2N+1) +2 b \big) e^{-\alpha \,x}+ (N\,\alpha +b)^2\,\right)
   \\ &
   = \ \frac{1}{2} \left(a^2\,z^2-a \,z\,\big(\alpha(2N+1) +2 b \big) + (N\,\alpha +b)^2\,\right) \ .
\end{aligned}
\end{equation}
The ground state function given by 
\begin{equation}
    \psi_0^{(N)} \ = \ e^{-\frac{a}{\alpha }\,e^{-\alpha  x}-b\,x }\,e^{-N\,\alpha \, x} \ = \ e^{-\frac{a}{\alpha }\,z+\frac{b}{\alpha}\,\log(z)}\,z^N  \ ,
\end{equation}
$P(z)=z^N$, having zero energy
\begin{equation}
    E_0^{(N)} \ = \ 0 \ .
\end{equation}
The corresponding SUSY partner potential obeys the relation
\begin{equation}
    V_1(x;N) \ =  \ V_0(x,N-1)\ ,
\label{}    
\end{equation}
hence, akin to the case of the exactly solvable Morse potential, in the QES case, $V_0$ and $V_1$ are shape invariant potentials; see in \cite{COOPER1995267} more details of the definition of shape invariant potential and the study of the exactly solvable Morse potential.

\section{Summary}

\vspace{0.2cm}

We studied two quasi-exactly solvable systems: the QES sextic and Morse potentials. For the sextic potential, approximate expressions for the WKB correction $\gamma$ and the energy spectrum $E_n(N)$ were constructed. Explicit analytical fits were given in the cases $N=0,~1/4,~1/2,~7/10$ where instanton effects are completely absent. At the level of algebraic operators which describe the polynomial QES solutions, the expression of the associated intertwining operator, in the variable $z=x^2$, were presented. Furthermore, a first-order SUSY transformation was performed to the QES Morse potential, showing that it is shape-invariant; as a consequence, the SUSY partner potential has the same hidden $\mathfrak{sl}_2(\mathbb{R})$ Lie algebraic structure with a shifted cohomology parameter $N \rightarrow N-1$. Generalizations of the exact WKB correction as well as the relation shape invariance- intertwiners  
\cite{Carrillo-Morales:2021ugo} for multidimensional systems (beyond separation of variables) is of great importance.
As a future analysis, it would be interesting to study higher-order SUSY partners of the QES Morse potential using excited states as seed functions. 
\vspace{0.2cm}

\section*{Acknowledgements}
ACA acknowledges Consejo Nacional de Humanidades Ciencia y Tecnolog\'{\i}a (CONAHCyT - M\'exico) support under the grant FORDECYT-PRONACES/61533/2020. A.M. Escobar Ruiz would like to thank the support from UAM research grant 2024-CPIR-0.

\section*{Data availability}
Data sharing is not applicable to this article as no new data were created or analyzed in this study. 

\bibliography{Biblio}


\end{document}